\documentclass[10pt]{iopart}

\usepackage{iopams}  
\usepackage{epsfig}
\usepackage{subfigure}
\usepackage{multirow}

\begin{document}

\title[Pentaquarks:  review of the experimental evidence]{Pentaquarks:  review of the experimental evidence}

\author{Sonia Kabana\dag\  
\footnote[3]{(sonja.kabana@cern.ch)}
}

\address{\dag\ Laboratory for High Energy Physics, University of Bern, Sidlerstrasse 5, 3012 Bern, Switzerland}

\begin{abstract}
Pentaquarks, namely baryons made by 4 quarks and one antiquark
have been predicted and searched for since several decades without success.
Theoretical and experimental advances in the last 2 years led
to the observation of a number of pentaquark candidates.
We review the experimental evidence for  pentaquarks
as well as their non-observations by some experiments,
and discuss to which extend these sometimes contradicting
informations may  lead to a consistent picture.

\end{abstract}




\section{Introduction}

\noindent
Pentaquarks is a name devoted to describe baryons made by 4 quarks and one antiquark.
Exotic pentaquarks have an antiquark with different flavour than their quarks e.g.  $uudd \overline{s}$.
For recent reviews on pentaquarks see \cite{jaffe_review,pdg}.
Pentaquarks have been predicted long 
time ago 
 e.g. \cite{pq_prediction,pra,diakonov_polyakov_petrov_1997}.
\noindent
Pentaquark searches were performed already in the 60'ies but few candidates
found have not been confirmed \cite{pdg_1986}.
However, recent significant advances in theoretical \cite{diakonov_polyakov_petrov_1997}
 and experimental work
 led to
 a number of new  candidates in the last 2 years of searches 
\cite{ nakano, clas_1, saphir,  clas_2, diana, hermes, neutrino, zeus, cosytof, camilleri,
 na49, na49_ximinus, h1, jamaica}.
\\

  \noindent
  Several models  predict the multiplet structure and characteristics of pentaquarks 
  for example the chiral soliton model, the uncorrelated quark model,  correlated
  quark models, QCD sum rules, thermal models,
  lattice QCD etc. (e.g. \cite{diakonov_polyakov_petrov_1997, navarra, jaffe, jaffe_2, octets_diakonov,ellis, guzey, stancu, glozman, aichelin_theta,stocker, 0402260, kopel}).
The current theoretical description of pentaquarks
 does not lead to a unique picture
on the pentaquark existence and characteristics, reflecting the complexity of the subject. 
  For example the observed mass splitting between $\Xi^{--}(1860)$ and
$\theta^+(1530)$ is unexpected or not allowed
by some authors
 like the old predictions of the soliton model
 \cite{karlinerlipkin,diakonov_polyakov_petrov_1997} and
expected by others 
 e.g. 
 new calculations of the soliton model
\cite{ellis,weigel}.
Furthermore, lattice calculations give very different results to the questions
if pentaquarks exist  and which mass and parity they have.

\noindent
In the following, we review the experimental observations of  pentaquark candidates,
as well as non-observations of candidates by some experiments.
We discuss to which extend these sometimes contradicting 
informations may   lead to a consistent picture.

\section{\bf Observations of pentaquark candidates}

\vspace{0.4cm}
\noindent
{\bf $\theta^+_{\overline{s}}$: $uudd \overline{s}$}

\noindent
 The first prediction
  of the mass of the state $uudd \overline{s}$ \cite{pra}
   using the chiral soliton model 
  was 
   m($uudd \overline{s}$) = 1530 MeV.
  A recent updated study within this model
  has shown that this value has a systematic error of the order of $\pm$ 100 MeV
  \cite{ellis}.

\begin{figure}[h]
\centering
\epsfig{figure=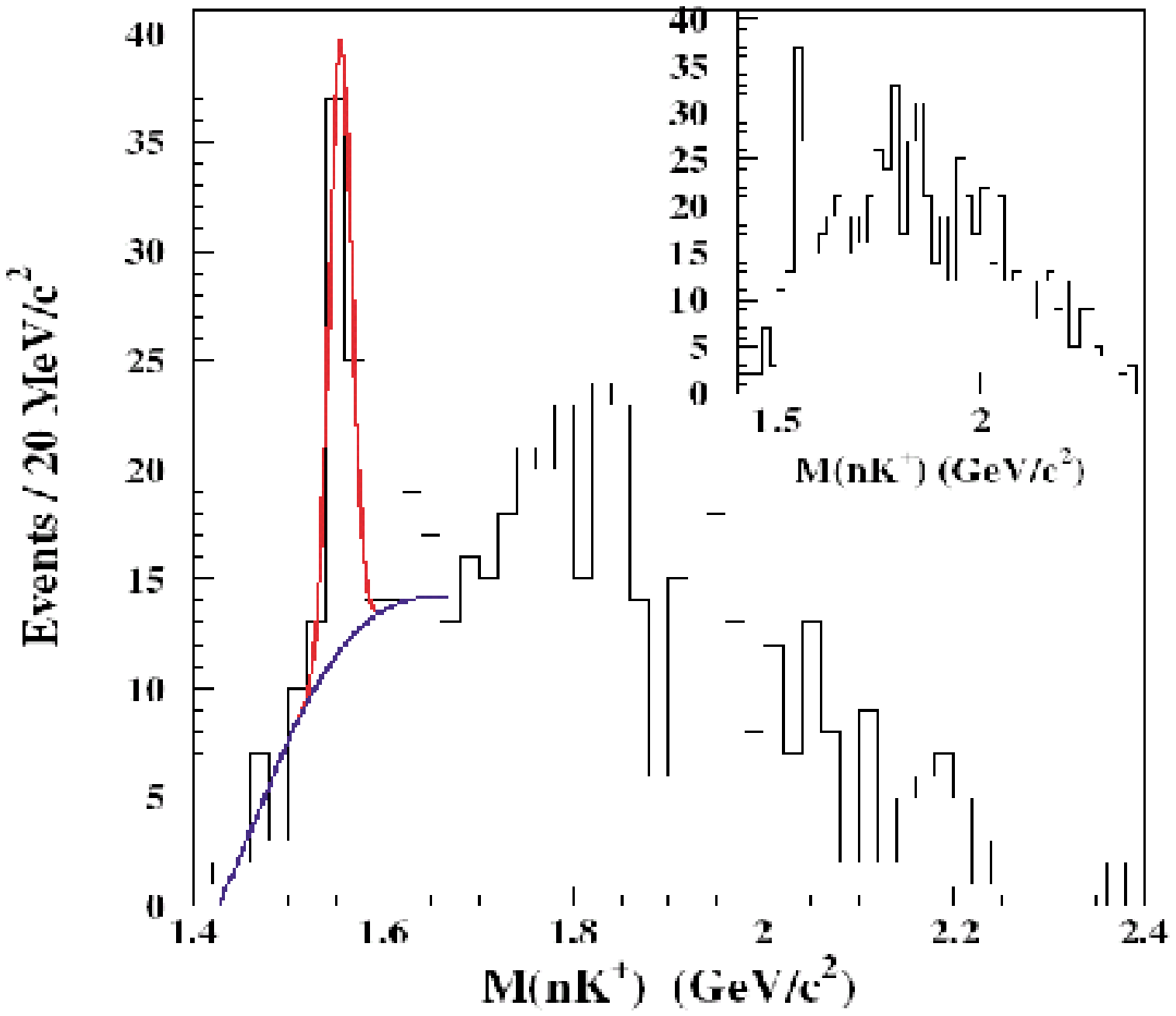,width=6cm}
\hspace*{0.4cm}
\epsfig{figure=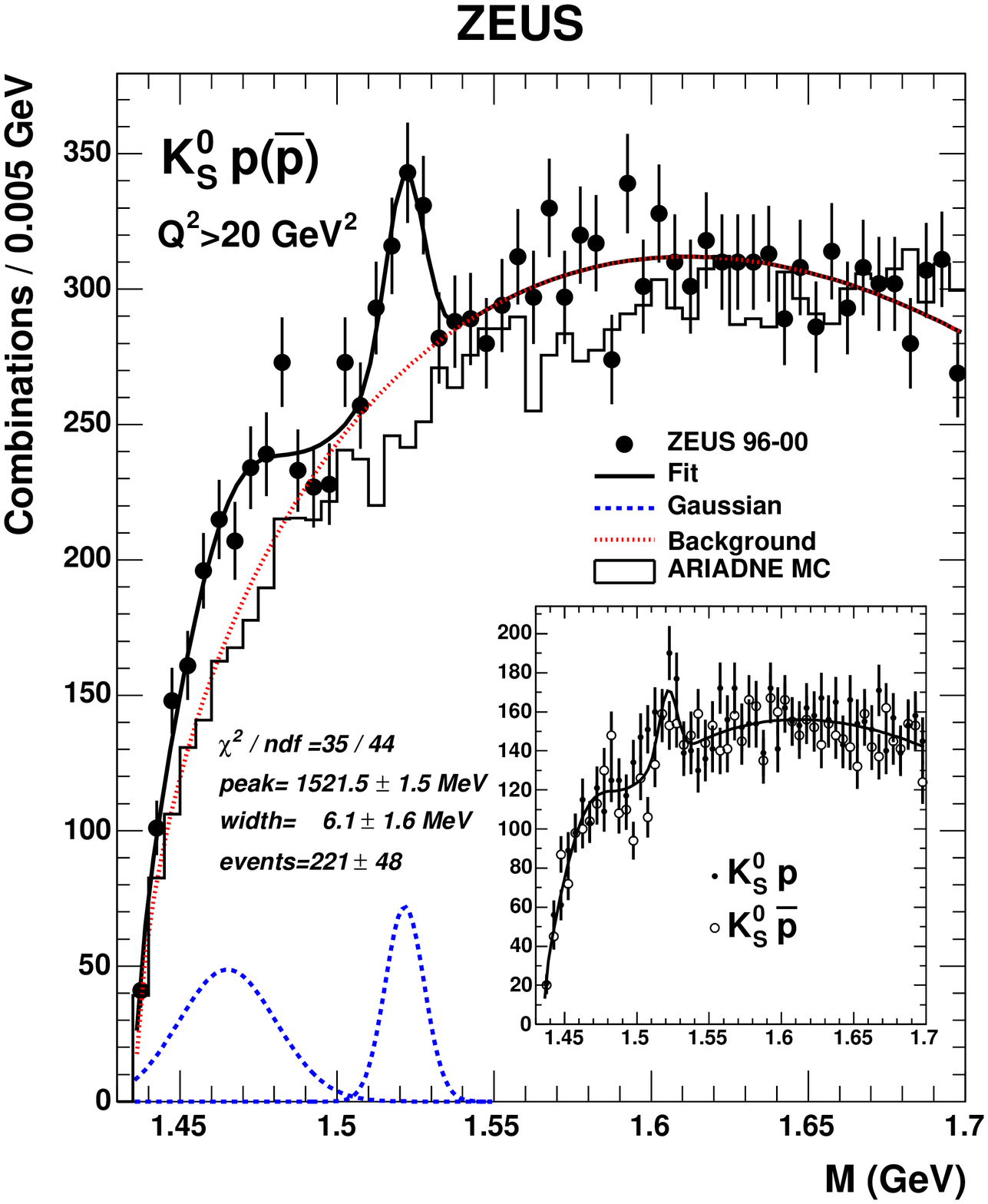,width=5cm}
\caption{Left: Invariant mass $n K^+$ measured by the CLAS Collaboration.
Right: Invariant mass $K^0_s p(\overline{p})$ measured by the Zeus Collaboration.}
\label{clas_zeus}
\end{figure}

 \noindent
  Recent
  advances in theoretical description of pentaquark characteristics,
  in particular the prediction of the width of $\theta^+$( $uudd \overline{s}$) 
  with spin=1/2 to be below 15 MeV 
  \cite{diakonov_polyakov_petrov_1997}
 as well as in experimental methods and instrumentation \cite{nakano}
  lead to
 the first observation of a candidate for the  $\theta^+$( $uudd \overline{s}$)
 in the $\gamma n \rightarrow  \theta^+ K^-   \rightarrow K^+ n K^- $
 reaction
 by the LEPS collaboration.
 They used $\gamma$ beam with energy 1.5-2.4 GeV.
 They found m( $\theta^+$) = 1540 $\pm$ 10 $\pm$ 5 (syst) MeV,
 width less than 25 MeV and $S/ \sqrt{B}$ = $19/\sqrt{17}$= 4.6.
 The
  neutron was inferred by missing mass measurement.
\\
Recent preliminary analysis of new data taken recently by LEPS lead to a confirmation
of the seen peak with about 90 entries in the peak above background, as compared
to 19 measured previously \cite{leps_pentaquark2004}.

\noindent
This first observation were followed by a number of experiments which have seen
the $\theta^+$ candidate peak.
The DIANA collaboration 
at ITEP (bubble chamber experiment)
  used $K^+$ beam with energy 850 MeV on
Xe and
have observed a peak in the invariant mass $K^0_s p$ from the
 reaction
$ K^+ Xe \rightarrow K^0_s p Xe^{'} $ 
\cite{diana}.
 They found m( $\theta^+$) = 1539 $\pm$ 2 MeV,
 width less than 9 MeV and $S/ \sqrt{B}$ = 4.4.

\begin{figure}[htb]
\centering
\epsfig{figure=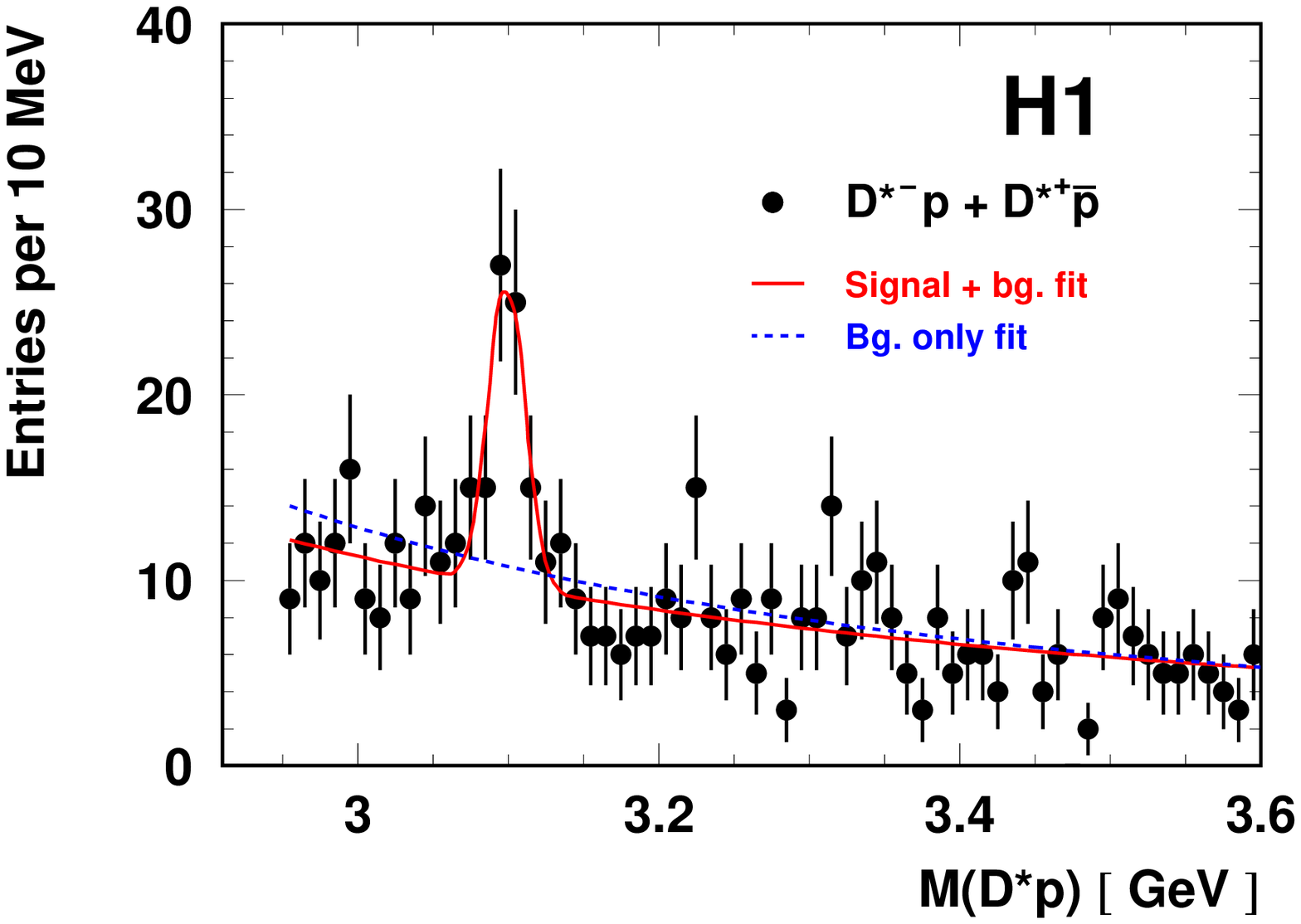,width=6.cm}
\epsfig{figure=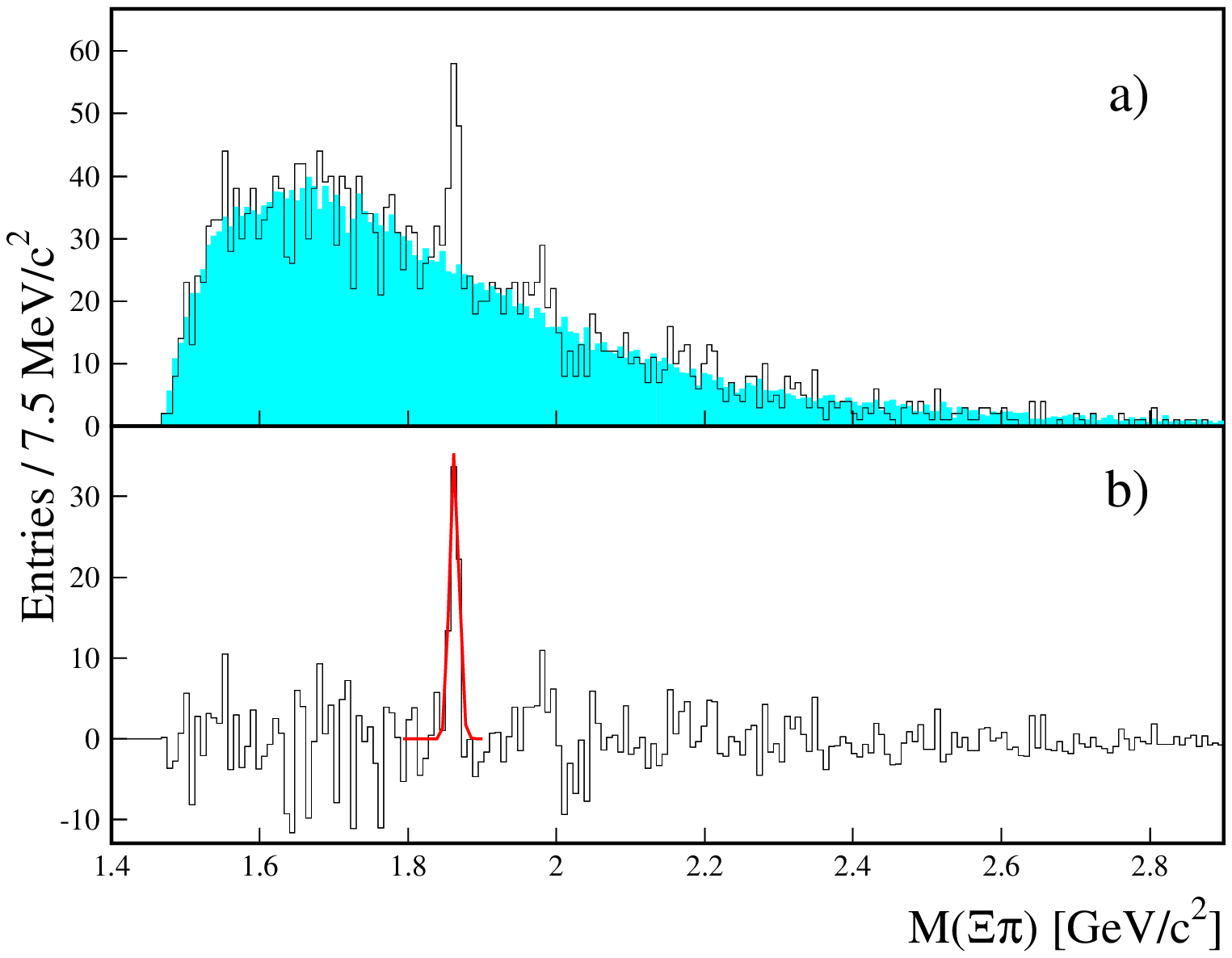,width=6.cm}
\caption{Left: Invariant mass $ D^{*-} p$ and $ D^{*+} \overline{p}$
measured by the H1 collaboration.
Right: Invariant mass of the sum of $\Xi^- \pi^-$, $\Xi^- \pi^+$ and their anti-channels
  measured by the NA49 collaboration.}
\label{h1_na49}
\end{figure}

\noindent
The SAPHIR  collaboration 
at ELSA
  used $\gamma$ beam with energies 31-94\% of 2.8 GeV on
H and
have observed a peak in the invariant mass $n K^+$ from the
 reaction
$ \gamma p \rightarrow \theta+ K^0_s  \rightarrow n K^+ K^0_s $ 
\cite{saphir}.
 They found m( $\theta^+$) = 1540 $\pm$ 4 $\pm$ 2 MeV,
 width less than 25 MeV and $S/ \sqrt{B}$ = 5.2.

\noindent
The HERMES  collaboration 
at DESY
  used $e^+$ beam with energy 27.6 GeV on
deuterium and
have observed a peak in the invariant mass $p K^0_s$
\cite{hermes}.
 They found m( $\theta^+$) = 1528 $\pm$ 2.6 $\pm$ 2.1 MeV,
 width 17 $\pm$ 9 $\pm$ 3 MeV and $S/ \sqrt{B}$ = 4.2 to 6.3.
While the signal to background ratio in the above publication is 1:3,
new analysis lead to an  improved signal to background ratio of 2:1
\cite{hicks_pentaquarks2004}.

\noindent
The COSY-TOF  collaboration 
 observed a peak in the invariant mass $p K^0_s$
 from the reaction $ p  p 
	  \rightarrow \Sigma^+ \theta^+
	  \rightarrow ( n \pi^+) ( K^0_s p)
    $
\cite{cosytof}.
 They found m( $\theta^+$) = 1530 $\pm$ 5 MeV,
 width below 18 $\pm$ 4 MeV and $S/ \sqrt{B}$ = 5.9.
They measure a cross section of 0.4 $\pm$ 0.1 $\pm$ 0.1 (syst) $\mu b$
which is in rough agreement with predictions of
0.1-1 $\mu b$
for p+p, p+n near threshold.

\noindent
An analysis of old $\nu$ and $\overline{\nu}$ interactions
from old bubble chamber experiments
filled with H, d, or neon,
and beam energies of 40 or 110 GeV
has resulted in a $\theta^+$ peak in the 
invariant mass $p K^0_s$
with
    m( $\theta^+$ ) = 1533 $\pm$ 5 MeV,
 width less than 20  MeV
and $S/ \sqrt{B}$= 6.7
\cite{neutrino}.

\noindent
The CLAS  collaboration 
  used $\gamma$ beam with energy about 95\%(2.474-3.115) GeV 
on deuteron target
and 
have observed a peak in the invariant mass $n K^+$
from the reaction 
$ \gamma d \rightarrow K^+ K^- p n 
$
through missing mass measurement of the neutron
\cite{clas_1}.
 They found    m( $\theta^+$) = 1542 $\pm$ 5 MeV,
 width of 21 MeV consistent with the experimental resolution,
  and $S/ \sqrt{B}$ = 5.2 $\pm$ 0.6.

\noindent
In a later publication the CLAS  collaboration 
  used $\gamma$ beam with energy 3-5.47 GeV 
on deuteron target
and 
have observed a peak in the invariant mass $n K^+$
from the reaction 
$ \gamma p \rightarrow \pi^+ K^- K^+ n 
$
through missing mass measurement of the neutron
 (fig. \ref{clas_zeus}, left)
\cite{clas_2}.
They observed evidence that the $\theta^+$ is preferably
produced through the decay of a new narrow resonance $N^0(2400)$.
 They found    m( $\theta^+$) = 1555 $\pm$ 10 MeV,
 width less than 26 MeV
  and $S/ \sqrt{B}$ = 7.8 $\pm$ 1.
This is the highest published significance obtained for the $\theta^+$ from
a single measurement.

\noindent
A preliminary analysis of CLAS
of the reaction
$\gamma d 
	\rightarrow \theta^+ \overline{ K^0} 
	\rightarrow (K^+ n) K^0_s$
 \cite{trento_bata}
  shows
two peaks in the invariant mass ($K^+ n$)
 at 1523 $\pm$ 5 MeV and at 1573 $\pm$ 5 MeV
both having a width of about 9 MeV
and significance of 4, respectively of 6 $\sigma$.
It is important that CLAS will clarify the reason for the shift of the
lower mass peak position and why the second peak appears with the new cuts
but not with the old ones in the previously studied reactions.
The second peak is a candidate for an excited $\theta^+$ state which is expected 
to exist with about $\sim$ 50 MeV higher mass than the ground state, in
agreement with the observation.
A preliminary cross section estimate gives 5-12 nb for the low mass peak
and 8-18  nb for the high mass peak.
\noindent
The above two peaks have been quoted also in \cite{bata_pentaquarks2004}.

\noindent
CLAS has taken a large amount of data in 2004 which are now been analysed.
First results have been quoted 
which confirm the previous $\theta^+$ observations with new peaks in different
channels, all near 1.55 GeV
 \cite{clas_pentaquarks2004}.

\noindent
The ZEUS  collaboration 
at DESY
  used $e^+p$ collisions at $\sqrt{s}$=300-318 GeV 
 and
have observed a peak in the invariant mass $p K^0_s$
 (fig. \ref{clas_zeus}, right)
  \cite{zeus}.
They have observed for the first time the $\overline{ \theta}^-$
state decaying in $\overline{p} K^0_s$.
 They found    m( $\theta^+$ + $ \overline{ \theta}^-$) = 1527 $\pm$ 2 MeV,
 width 10 $\pm$ 2 MeV.

\noindent
The NA49 experiment has also reported recently
a preliminary result of a peak observed in the invariant mass $p K^0_s$ 
in p+p reactions at $\sqrt{s}$=17 GeV
with mass 1526 $\pm$ 2 MeV and width below 15 MeV
\cite{kadija_pentaquark2004}.
They have also reported a preliminary evidence that the
$\theta^+$ peak appears more pronouned after
assuming $\theta^+$ production through the decay of a resonance
$N^0(2400) \rightarrow \theta^+ K^-$ \cite{barna}
as suggested by CLAS \cite{clas_2} and discussed in
\cite{karliner,azimov_n2400}.

\noindent
NOMAD has shown recently 
preliminary results on the observation of a 
$\theta^+$ candidate peak in the $p K^0_s$ invariant mass
using their full statistics of $\nu A$ interactions, with a mean energy of 
the $\nu$ beam of 24.3 GeV \cite{camilleri}.
The mass observed is 1528.7 $\pm$ 2.5 MeV
and the width is consistent with the 
experimental resolution of 9 MeV.

\noindent
GRAAL has shown preliminary results on the observation of a 
$\theta^+$ candidate peak in the $p K^0_s$ invariant mass
in 
 $\gamma d \rightarrow \theta^+ \Lambda^0 \rightarrow (K^0_s p) \Lambda^0$
 interactions, using $\gamma$ energy of maximally 1.5 GeV \cite{graal_pentaquark2004}.
The mass observed is 1531 MeV while no error is given.
\\

\noindent
Most of the experiments measure a $\theta^+$ width consistent with the experimental resolution,
while two of them give a measurement of width somewhat larger than their resolution
namely Zeus and Hermes.
A measurement with a much improved resolution would be important.
\\
Non-observation of $\theta^+$ in previous experiments lead to an estimate
of its width to be of the order of 1 MeV or less \cite{1mev_arndt}.
This limit would gain in significance, once the $\theta^+$
non-observation by several experiments will be better understood,
excluding other reasons for the $\theta^+$ non-observation in the examined
reactions.

\vspace*{0.4cm}
\noindent
{\bf Do the measured $\theta^+$ masses vary as expected for a real state ?}

\begin{figure}[htb]
\centering
\epsfig{figure=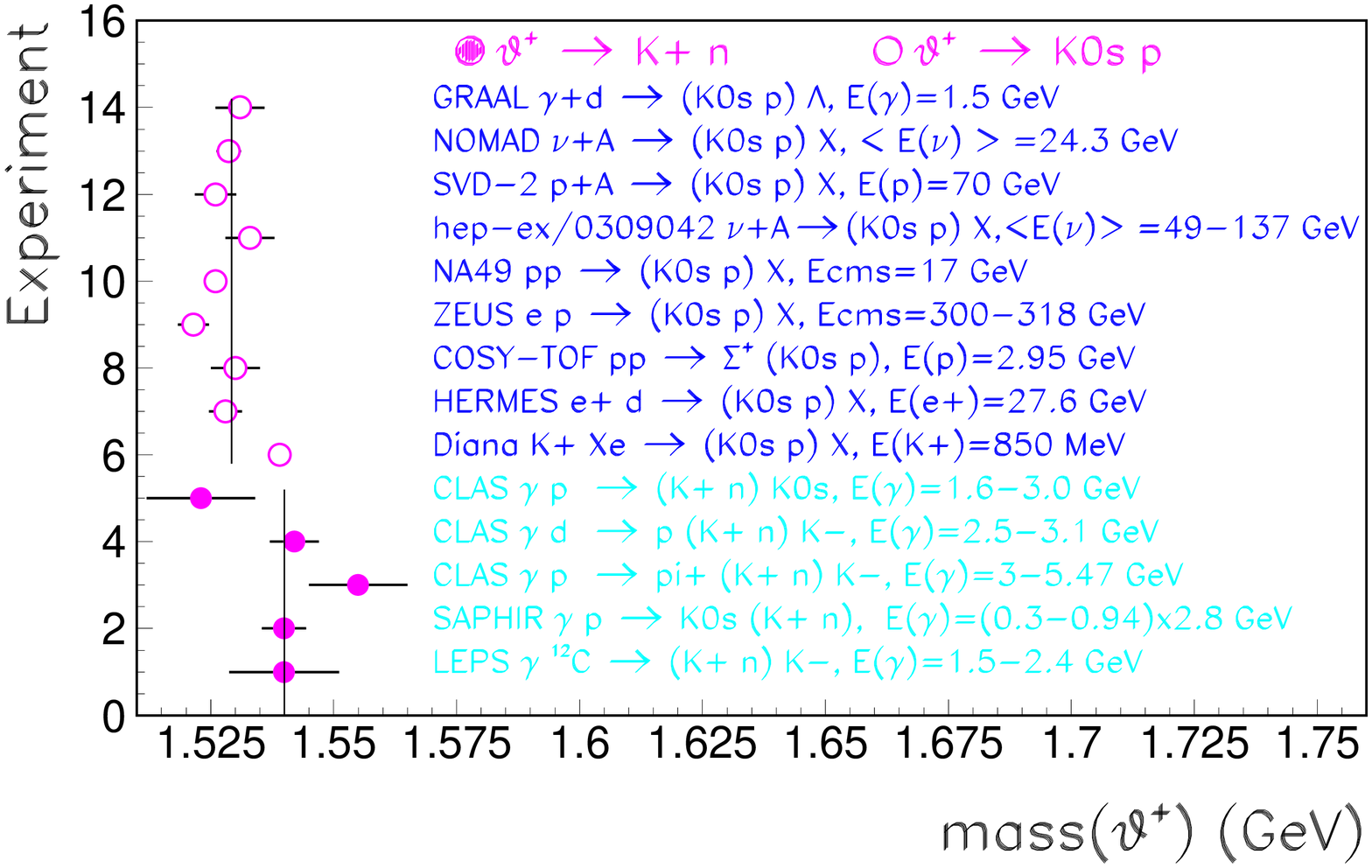,width=12cm}
\caption{Compilation of measured masses of $\theta^+$ candidates.}
\label{theta_mass}
\end{figure}

\noindent
Figure \ref{theta_mass} shows a compilation of the 
masses of $\theta^+$ candidate peaks observed
by several experiments.
The statistical and systematic errors (when given) have been added in quadrature.
For GRAAL 
we assume an error of 5 MeV as no error has been given in \cite{graal_pentaquark2004}.
For the two preliminary peaks of CLAS we assume the systematic error
of 10 MeV quoted previously by CLAS.
The lines indicate the mean value of the mass among 
the $\theta^+ \rightarrow p K^0_s$
and the
$\theta^+ \rightarrow n K^+$
observations.
It appears that the 
mass of $\theta^+$
from $\theta^+ \rightarrow n K^+$
observations
is systematically higher than the one
from 
$\theta^+ \rightarrow p K^0_s$
observations.
This may be related to the special corrections needed
for the Fermi motion and/or to details of the
analysis with missing mass instead of direct measurement 
of the decay products.

\noindent
All observations together give a mean mass of 
1.533 $\pm$ 0.023 GeV and they
deviate from their mean with a
 $\chi^2/DOF$
of 3.92.
The $\chi^2/DOF$ for 
the deviation of the $\theta^+ \rightarrow p K^0_s$
observations
from their mean of 1.529 $\pm$ 0.011 GeV
is 
3.76.
The $\chi^2/DOF$ for 
the deviation of the $\theta^+ \rightarrow n K^+$
observations
from their mean of 1.540 $\pm$ 0.020 GeV
is 0.94.

\noindent
The bad $\chi^2/DOF$ for the $\theta^+ \rightarrow p K^0_s$
observations
maybe due to an underestimation of the systematic errors.
In particular in some cases no systematic errors are given,
sometimes because the results are preliminary.
If we add a systematic error of 0.5\% of the measured mass (therefore
of about 8 MeV) on all
measurements for which  no systematic error 
was given by the experiments,
we arrive to a
$\chi^2/DOF$ for the $\theta^+ \rightarrow p K^0_s$
observations
of 0.95
and a mean mass of 1.529 $\pm$ 0.022 GeV.
The $\chi^2/DOF$ for the $\theta^+ \rightarrow n K^+$
observations
almost don't change by this,
(mean mass = 1.540 $\pm$ 0.022 GeV, $\chi^2/DOF$=0.91),
because the experiments mostly give the systematic errors
for this decay channel.
All observations together give then a mean mass of 
1.533 $\pm$ 0.031 GeV and they
deviate from their mean with a
 $\chi^2/DOF$
of 2.1, reflecting mainly the difference of masses between
the two considered decay channels.
It is important to understand the origin of this discrepancy.
This
 can be studied measuring $\theta^+ \rightarrow K^+ n$ in experiments
with direct detection of the neutron or the antineutron for the $\overline{ \theta^-}$
 like PHENIX \cite{phenix} and GRAAL.

\vspace*{0.4cm}
\noindent
{\bf $\theta^{++}$ }

\noindent
A preliminary peak is quoted by CLAS \cite{bata_pentaquarks2004}
 for the candidate $\theta^{++} \rightarrow
p K^+$ produced in the reaction 
$ \gamma p \rightarrow \theta^{++} K^-
		\rightarrow p K^+ K^-$
 at 1579 $\pm$ 5 MeV.
A previous peak observed by CLAS in the invariant mass $ p K^+$
has been dismissed as due to $\phi$ and hyperon resonance reflexion
\cite{clas_thetapp_1}.
\\
\noindent
The STAR collaboration quoted a preliminary peak in the $p K^+$ and 
$\overline{p} K^-$ invariant
masses at 1.530 GeV, with $S/\sqrt{B}$ $\sim$ 3.8 and width about 9 MeV,
  which is candidate
for the
 $\theta^{++} \rightarrow p K^+$
as well as the antiparticle
 $\overline{ \theta^{--}}  \rightarrow \overline{p} K^-$
in d+Au collisions at $\sqrt{s}$=200 GeV \cite{star_thetapp}.

\vspace*{0.4cm} 
\noindent
{\bf $\Xi$, $N^0$}

\noindent
The NA49 experiment has observed in p+p reactions
at $\sqrt{s}$=17 GeV 
the pentaquark candidates 
$\Xi^{--}(1862 \pm 2  MeV) \rightarrow \Xi^- \pi^-$,
the
$\Xi^{0}(1864 \pm 5  MeV) \rightarrow \Xi^- \pi^+$
and their antiparticles  \cite{na49}
(figure \ref{h1_na49}, right).
They measure a width consistent with their resolution of about  18 MeV.
They also observe preliminary results of the decay
$\Xi^-(1850) \rightarrow \Xi^0(1530) \pi^-$
with simarly narrow width as the other candidates \cite{na49_ximinus}.
The $\Xi^{--}$ is a candidate for the antidecuplet 
and the $\Xi^0$ too due to the small mass difference
while it is unclear if the $\Xi^-$(1850) is 
from the octet or the antidecuplet.
An observation of the $\Xi$ I=1/2 from the octet in $\Lambda K^0_s$
would answer this question.
 The non observation of a peak in the invariant mass $\Xi^0(1530) \pi^+$
 is also an important information, as this decay channel
 is not allowed by SU(3)
 for the antidecuplet $\Xi^+$ \cite{jaffe_2}.

\begin{figure}[htb]
\centering
\epsfig{figure=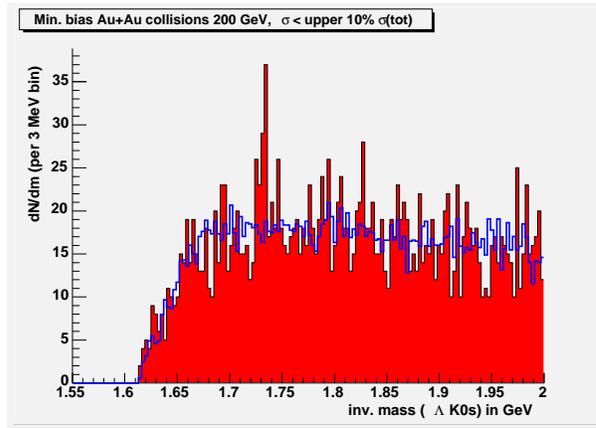,width=8.cm}
\caption{Invariant mass $\Lambda K^0_s$ measured by the STAR experiment.}
\label{star}
\end{figure}

\noindent
The experiment STAR has shown preliminary results on a 
$N^0$ ($udsd \overline{s}$, $udd u \overline{u}$)
 or $\Xi$ ($uds s \overline{d}$) I=1/2 candidate \cite{jamaica}.
STAR uses minimum bias Au+Au collisions at $\sqrt{s}$=200 geV
and  observes a peak in the decay channel
	$\Lambda K^0_s$ at a mass 1734 $\pm$ 0.5 (stat) $\pm$ 5 (syst) MeV
 (fig. \ref{star})
with width consistent with the experimental resolution of about  6 MeV
and $S/ \sqrt{B}$ between 3 and 6 depending on the method used
\cite{jamaica}.
They don't observe a peak near 1850 - 1860 MeV 
 resulting from a $\Xi^0$ I=1/2 (octet) pentaquark state
with the same mass as the $\Xi^-(1850)$ of NA49,
disfavouring the picture of degenerate octet and antidecuplet 
even though a low branching ratio to $\Lambda K^0_s$
may not allow to observe the peak with the present statistics.

\noindent
The GRAAL experiment has shown preliminary results on 
two narrow  $N^0$ candidates.
One candidate is observed at a mass of 1670 MeV in  the invariant mass of
$ \eta n$ from the reaction
$ \gamma d \rightarrow \eta n X$.
 The neutron has been directly detected.
The other is observed at a mass of 1727 MeV
 in the invariant masses of 
$\Lambda K^0_s$  as well as 
in the invariant masses of
$\Sigma^- K^+$
at the same mass and with the same width
\cite{trento_graal}.
The second reaction allows to establish the strange quark content  and
therefore to exclude the $\Xi$ hypothesis.
The difference of 7 MeV between the STAR and GRAAL 
measured masses of 1727 and 1734 MeV, should be compared to the
systematic errors. STAR quotes a systematic error of 5 MeV
while GRAAL quotes no systematic error.

\noindent
The mass of the peaks  at 1670 and at  (1727,1734) MeV  is in  good agreement with the
$N$ masses 
  suggested by Arndt et al \cite{arndt0312126}. In this paper
a modified Partial Wave analysis allows to search for narrow
states and presents two candidate $N$ masses, 1680 and/or 1730 MeV
with width below 30 MeV.

\noindent
While the above mentioned narrow $N^0$ candidates 
of mass 1670 and 1727-1734 MeV
fit well into the picture of the expected $N$ and $N_s$ pentaquark
candidates, they can also be something else than pentaquarks,
 e.g. a new $N^0$=$udd$ resonance, or a $udd gg$ state.

\vspace{0.4cm} \noindent 
{\bf $\theta^0_{\overline{c}}$}

\noindent
The H1  collaboration 
at DESY
  used $e^- p$ collisions at $\sqrt{s}$=300 and 320 GeV 
 and
have observed a peak in the invariant masses
$ D^{*-} p$
and
$ D^{*+} \overline{p}$
(figure \ref{h1_na49}, left)
at a mass 3099 $\pm$ 3 (stat) $\pm$ 5 (syst) MeV
and width of 12 $\pm$ 3 MeV  \cite{h1}.
The probability that the peak is a background fluctuation is less than $4 \ 10^{-8}$.
Extensive systematic studies have been performed.
The momentum distribution of the signal is as expected for a particle at this mass.
This peak is a candidate for the state $\theta^0_{\overline{c}} $ = $uudd \overline{c}$
and is the first charmed pentaquark candidate seen.
The mass and width of the particle and it's antiparticle are consistent.

\vspace{0.4cm} \noindent
\section{\bf Lack of observation of pentaquark candidates}

\noindent
Several experiments have reported preliminary or final results on the non-observation
of pentaquarks e.g. $e^+e^-$: Babar, Belle, Bes, LEP experiments, 
$p \overline{p}$: CDF, D0 $pA$:E690, $eA$: HERA-B, $ep$: Zeus (for the
$\theta^0_c$) \cite{non}.
 HERMES has reported the non-observation of a peak 
 in the $p K^+$ invariant masses \cite{hermes}.
\\
It has been  argued that the non-observation of pentaquark
states in the above
experiments could be due to an additional strong suppression factor
for pentaquark production
in $e^+ e^-$ collisions, as well as in B decays 
which is lifted in reactions like $\gamma A$
 in which a baryon is present in the initial state
\cite{karliner}.
The constituents of the $\theta^+$ are already present in the initial
state of e.g.  low energy photoproduction experiments, while
in other experiments baryon number and strangeness must be created from gluons \cite{karliner}.
It is important to try to assess the expected cross sections.
\\
The CDF ($p \overline{p}$) and E690 (pA) non observation of pentaquarks
can be a consequence of the decrease of the pentaquark cross section
with increasing energy \cite{0406043,titov}.
This depends however on the kinematic region considered, and 
it is suggested to look for pentaquarks in the central rapidity region  \cite{0406043,titov}.
\\
In addition, if the $\theta^+$
is produced preferably through the decay of a new resonance 
$N^0(2400) \rightarrow \theta^+ K^-$ 
as suggested by CLAS and NA49 and as discussed in \cite{karliner,azimov_n2400},
neglecting this aspect maybe a further cause of its non-observation in some experiments.
\\
Some authors point out the importance to 
exclude kinematic reflexions as reason behind the $\theta^+$ peak
\cite{dzierba}.
This known source of systematic errors is investigated by the experiments
which observe pentaquark candidates.
Other authors discuss limits from the non observation
of the $\Xi(1860)$ peaks of NA49 by previous experiments \cite{wenig}.
\\

\vspace*{0.2cm}
\noindent
It is clear that a higher statistic is desirable in order to confirm
the pentaquark observations reported so far, as well as more measurements
and searches by other experiments.
New data taken in 2004 and planned to be taken in 2005 will
lead to enhancements in statistics of experiments up to a factor of 15
allowing to test the statistical significance and make more systematic studies.
Experiments searching for pentaquarks should  test also the production mechanisms 
proposed in the literature e.g. the $\theta^+$ production
through the $N^0(2400)$ decay.
For example
 Phenix could  search for the final state
$ \overline{ \theta^-} K^+ $
or 
$ \overline{ \theta^-} K^0_s $
demanding the invariant mass of $ \overline{ \theta^-} K^0_s $ and
$ \overline{ \theta^-} K^+ $
to be in the range 2.3 to 2.5 GeV,
and study the option to trigger online on this channel.

\section{Summary and conclusions}

\noindent
Recent theoretical and experimental advances led to the observation of
candidates for a number of pentaquarks states.
In particular candidate signals have been observed for the
$\theta^+_{\overline{s}}(1533)$, $\theta^{++}_{\overline{s}}$(1530/1579),
$\theta^0_{\overline{c}}$(3099),
$\Xi^{--}(1862)$,
$\Xi^{0}(1864)$,
$\Xi^{-}(1850)$,
$N/\Xi^0(1734/1727)$,
$N^0(2400)$,
$N^0(1670)$ states
as well as a possible excited $\theta^+_{\overline{s}}(1573)$ state.
These
 observations are very promising, however despite the large number of e.g. the
$\theta^+_{\overline{s}}$ observations, they all suffer
 from a low individual statistical significance and this is what really matters.
A 
much higher statistics is needed to support and solidify the existing evidence.
Several other experiments have reported lack of observation of those candidates.
The
 inconsistency among experiments waits to be clarified through
carefull high statistics measurements of pentaquark candidates, their
characteristics (cross sections, quantum numbers)  and  upper limits in the case
of non-observation. Furthermore, systematic studies are needed as well as advanced theoretical
understanding of the observations and the reasons behind non-observations. 
Combined theoretical and experimental efforts
should resolve the current puzzle and 
answer the question if we have discovered a new type of baryons, the pentaquarks.

\end{document}